\documentclass[%
 reprint,
 superscriptaddress,
nofootinbib,
 amsmath,amssymb,
 aps,
]{revtex4-2}

\usepackage{xcolor}
\usepackage{graphicx, cleveref, siunitx, subfigure} % Include figure files
\usepackage{dcolumn}% Align table columns on decimal point
\usepackage{bm}% bold math
\begin{document}

\preprint{APS/123-QED}

\title{Speeding axion haloscope experiments  using heterodyne-variance-based detection with a power-meter}

\author{Zhanibek Omarov}
\affiliation{Department of Physics, KAIST, Daejeon 34141, Republic of Korea}
\affiliation{Center for Axion and Precision Physics Research, IBS, Daejeon 34051, Republic of Korea}
\author{Junu Jeong}
\email{jwpc0120@gmail.com}
\thanks{Corresponding author}
\affiliation{Center for Axion and Precision Physics Research, IBS, Daejeon 34051, Republic of Korea}
\author{Yannis K. Semertzidis}
\email{semertzidisy@gmail.com}
\thanks{Corresponding author}
\affiliation{Center for Axion and Precision Physics Research, IBS, Daejeon 34051, Republic of Korea}
\affiliation{Department of Physics, KAIST, Daejeon 34141, Republic of Korea}

\date{\today}% It is always \today, today,
\bibliographystyle{apsrev4-2}

\begin{abstract}
We describe a new axion search method based on measuring the variance in the interference of the axion signal using injected photons with a power detector.
The need for a linear amplifier is eliminated by putting a strong signal into the microwave cavity, to acquire not only the power excess but also measure the variance of the output power.
The interference of the external photons with the axion to photon converted signal greatly enhances the variance at the particular axion frequency, providing evidence of its existence. 
This method has an advantage in that it can always obtain sensitivity near the quantum noise limit even for a power detector with high dark count rate.
We describe the basic concept of this method both analytically and numerically, and we show experimental results using a simple demonstration circuit. 
\end{abstract}

\maketitle

%\tableofcontents

\section{Introduction}

The axion particle, which is theoretical evidence that the strong CP problem has been solved with spontaneous symmetry breaking of a global chiral symmetry~\cite{PRL1977PQ,PRL1978Weinberg,PRL1978Wilczek}, is also a good candidate for dark matter when it has a very light mass~\cite{PLB1983Wilczek,PLB1983Abbott,PLB1983Dine} through the KSVZ~\cite{PRL1979Kim,NPB1980SVZ} or the DFSZ models~\cite{YF1980Zhitnitsky,PLB1981DFS}.
The haloscope method that converts dark matter axions into microwave photons, using a cavity resonance in strong magnetic fields, as proposed by Pierre Sikivie~\cite{PRD1985Sikivie}, is still the most sensitive method to search for invisible axions.
The total power of axion to microwave photon conversion rate is:
\begin{equation}
    P_{\rm conv.} = \frac{g_{a\gamma\gamma}^{2}\rho_{a}}{m_{a}}\langle \mathbf{B}^{2} \rangle V_{c} C \frac{Q_{c}Q_{a}}{Q_{c} + Q_{a}},
\end{equation}
where $g_{a\gamma\gamma}$ is the axion-photon coupling, $\rho_{a}$ is the local dark matter axion density, $m_{a}$  the axion mass, and $\langle \mathbf{B}^{2} \rangle$ is the average of the square of the applied magnetic field over the cavity volume $V_{c}$.
$C$ is the form factor describing how well the resonant mode is matched with the applied magnetic field while $Q_{c}$ and $Q_{a}$ are the quality factors of the cavity and axion, respectively~\cite{JCAP2020Kim}.
The axion line-shape was estimated by Turner~\cite{article:Turner} to be a Lorentzian, with a quality factor approximated to be about $10^6$.

For example, under an average magnetic field of about $10\,$T, a DFSZ axion dark matter radiates $10^2\,$RF-photons/s in a $37\,$liter cavity with a quality factor of $10^5$, and around 20\% of them are transmitted to a receiver circuit when an antenna is optimally inserted into the cavity with a coupling of 2~\cite{JCAP2020Kim}.
The thermal photons from the cavity that couple to the outside circuit are a primary source of noise, $50\,$mK cavity temperature with an integration time of $1\,$s corresponds to the same equivalent power level with a S/N ratio of 1, according to the Dicke radiation formula~\cite{RSI1946Dicke}.
However, the amplifiers needed to enhance the available power level add their own noise; even for the best linear RF-amplifiers the added noise is many times the power expected from the axion conversion.
In addition, linear amplifiers have an unavoidable minimum noise limit due to quantum fluctuations of the Heisenberg uncertainty principle depending on the target frequency $f_{c}$: $48\,{\rm mK}\times (f_{c} / 1\,{\rm GHz})$.
The latest quantum amplifiers typically add noise in the range of 1 to 10 times higher than this quantum limit~\cite{NPhysics2012HoEom,PRL2017Brubaker,PRXQ2021Malnou,SupercondSci2021Kutlu,PRL2021ADMX,PRL2021Kwon}.

The axion search can essentially be reduced by trying to reveal its frequency location amongst the potentially millions of frequency channels it could be hiding in.
If the axion frequency is known accurately enough, then an experiment with reasonable sensitivity could determine whether axions constitute the local dark matter or not in less than a day of integration time, even if they were only a fraction of the local dark matter halo density.
To find the axion frequency, experiments scan for the possible axion masses aiming to observe an anomalous photon excess.
In this process, the phase information of the dark matter axion is discarded.
However, using the knowledge that the phase of the axion is constant for an estimated coherence time, it is possible to easily reach the quantum limit by constructing an interferometer with precisely controlled external photons and measuring the interference effect at every coherence interval.

There has been impressive progress within the last decade towards theoretically interesting sensitivities in the axion dark matter field~\cite{PRL2021ADMX,PRL2021Kwon,Nature2021Backes,SciAdv2022Yannis}.
Nonetheless, since the possible axion frequency range is vast, the field can use more sensitive and simplified methods to boost the scanning rate by a factor of a few to an order of magnitude.
In this paper we propose a new method to look for the axion signal impended in a thermal noise by injecting probe photons into the cavity from the outside and interfering them with the axion induced photons.
This method is similar to the heterodyne detection developed~\cite{Bush2019PRD} for ALPs at DESY, see also~\cite{arXiv2021Sikivie}, with a significant difference: that in the axion dark matter case the axion phase is constant only within the axion coherence time and hence detecting the variance  is more appropriate.
The method is advantageous to the traditional detection method when the linear amplifier noise temperature is more than about one photon quanta.

\section{Heterodyne Haloscope}

Heterodyne detection is a signal processing technique frequently used in microwave engineering, and it is used to raise or lower the frequency of a signal by interfering it with a coherent reference photon of known frequency, using a mixer.
In optics, since the observer itself is a power-detector, not a field detector, and it has a ``square-law'' characteristic, interference effects naturally appear when several frequency components, signal and probe, enter together.
Heterodyne detection is an experimental technique often used in optics for the following purposes:
(1) amplifying a weak signal, (2) detecting a phase change in a signal, (3) lowering the frequency through beating, and (4) reducing the noise contribution of a power detector.

For effective interference, the heterodyne detection technology must maintain the coherence of the two frequencies~\cite{Bush2019PRD}.
In the case of a local oscillator at microwave frequencies, coherence can be maintained with very high precision.
However, the photon signal induced from the dark matter axion field cannot maintain long-time coherence, because it is limited by its quality factor.
This implies constructive and destructive interferences are repeated at regular (but unknown) periods in the interferometer so that the average power excess will be very close to zero.

However, the expected quality factor of the axion dark matter around the solar system can be obtained by the virial theorem~\cite{article:Turner}.
Since the relative phase of the dark matter axion to the local oscillator for every interference is given arbitrarily, it is also arbitrary whether the interference at every coherence time interval is constructive or destructive.
Therefore, the variance in the interference for each coherence time interval can reveal the existence of the axion using the advantages of a heterodyne interferometer.
A mathematical description of the amplification of the variance of the axion signal by interfering with the probe is described in appendix~\ref{sec:var_enhance}.

\begin{figure}[t]
  \centering
  \includegraphics[width=0.99\linewidth]{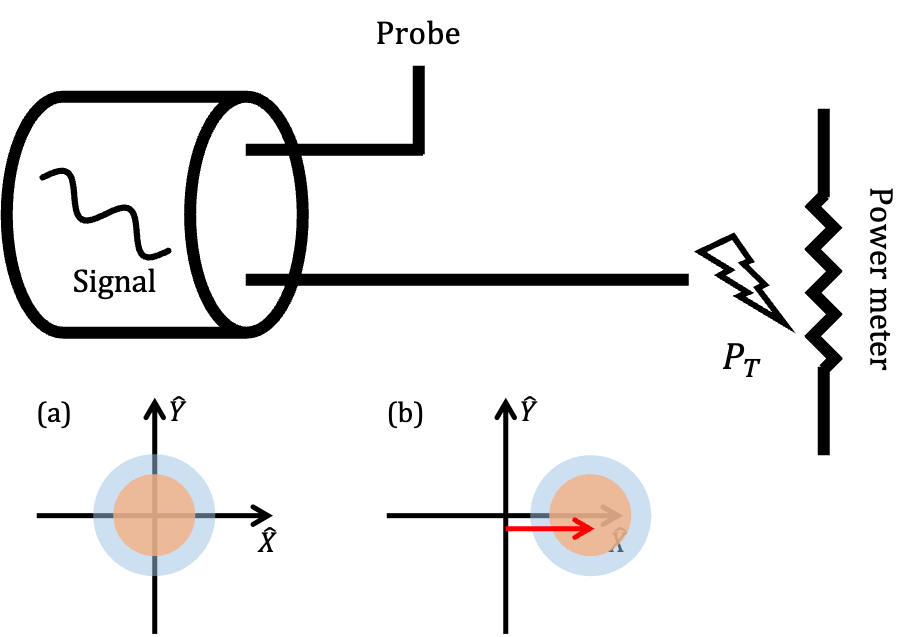}
  \caption{Detector schematics for a variance measurement setup and in-phase and quadrature components (a) without a probe and (b) with a probe.
  The orange and the blue circles refers to the noise and the signal, respectively, and the red arrow indicates the probe case.
  }\label{fig:schematics}
\end{figure}

Assuming that the axion and the probe signals are sine waves with the same frequency and random relative phases, the superposition of such waves is either constructive or destructive.
Although the random relative phase suggests that the average effect is nullified, the variance of such repetitive superposition is always finite.
\Cref{fig:schematics} shows a schematic setup and the underlying procedure in the in-phase and quadrature (IQ) frame for such an axion detection method.
Injecting the probe signal simply shifts the signal and the noise in the IQ plane, so it doesn't enhance the signal-to-noise ratio when a field measurement approach is applied.
In the case of power detectors, however, the probe increases the variance of the detected photons, signal and noise, while the dark count of the detector is unchanged.
A simple experiment showing the variance enhancement through the application of probe microwave-photons was performed and is briefly described in appendix~\ref{sec:test}.
Therefore, the variance detection with a probe can enhance the signal-to-noise ratio when the dark count of a power detector is the dominant noise source, which is the case for most current microwave bolometers~\cite{Nature2019Roope,Nature2020Lee}.

When a candidate axion signal location is found, it can also be marked for cross-checking by re-scanning using the most sensitive amplifier available at the time, and applying the traditional power-level detection method for the particular frequency channel. 

\section{Amplified-variance detector sensitivity}

The number of transmitted photons on a power-meter per sampling, $N_{T}$, is the sum of the signal photons, $N_{s}$, the probe photons, $N_{p}$, and the noise photons, $N_{n}=N_{\rm th.} + N_{D}$ where $N_{\rm th.}$ corresponds to the thermal noise photons and $N_{D}$ is the dark count photons.
The variance of the transmitted photons for each sample, for a signal with random phase,  is:
\begin{equation}
\label{eq:variance_photon}
    \hat{\sigma}^{2}(N_{T}) = N_{\rm th.} + N_{D} + N_{s} + N_{p}\left( 1 + N_{s}+N_{\rm th.} \right),
\end{equation}
where $\hat{\sigma}^{2}$ is the variance estimator.
The first three terms are a pure variance of the noise and the signal without the probe, and the last term is the probe-amplified variance.
The signal-to-noise ratio is defined by the excess divided by the error of the estimator:
\begin{equation}
\label{eq:snr}
    \begin{split}
        {\rm S/N}_{\hat{\sigma}^{2}} \equiv & \frac{\hat{\sigma}^{2}(N_{T}) - \hat{\sigma}^{2}(N_{T})|_{N_{s} = 0}}{ \sqrt{ \hat{\sigma}^{2} (\hat{\sigma}^{2}(N_{T})|_{N_{s} = 0}) }} \\
        \approx& \frac{N_{s} + N_{s} N_{p}}{ (N_{D} + N_{p}) \sqrt{ 2 + 1 / (N_{D} + N_{p}) + \mathcal{O}(N_{\rm th.}) } } \sqrt{n},
    \end{split}
\end{equation}
where $n$ is the number of samples for the variance measurement.
When deriving the last line, we assumed a sufficiently large $n$ and a very small $N_{\rm th.}$.
A detailed derivation is described in Appendix \ref{sec:snr}.

For cryogenic temperatures, the bosonic occupation number of the thermal photons decreases exponentially and the zero-point fluctuations can be neglected via a single quadrature measurement.
In other words, for adequately low temperatures, this implies that $N_{\rm th.} \to 0$.
Therefore, in the absence of external driving power, i.e. $N_{p} = 0$, the S/N of variance detection is proportional to the ratio of the signal counts to the dark counts (${\rm S/N}_{\hat{\sigma}^{2}} \propto N_{s} / N_{D}$ for $N_{D} \gg 1$ and ${\rm S/N}_{\hat{\sigma}^{2}} \propto N_{s} / \sqrt{N_{D}}$ for $N_{D} \ll 1$).
On the other hand, when there is a sufficient amount of probe photons, i.e., $N_{p} \gg 1$, the dark count elements in the S/N disappear and the S/N becomes proportional only to the number of signal photons.

For a detector with a sampling rate of $f_{s}$, the number of photons per sampling is the ratio of the photon rate to the sampling rate.
So the S/N for the photon rate is:
\begin{equation}
\label{eq:SNR_rate}
\begin{split}
    {\rm S/N}_{\hat{\sigma}^{2}} \approx & \frac{\dot{N}_{s} + \dot{N}_{s} \dot{N}_{p} /  f_{s} }{ (\dot{N}_{D} + \dot{N}_{p}) \sqrt{ 2 + f_{s} / (\dot{N}_{D} + \dot{N}_{p}) } } \sqrt{f_{s} \Delta t} \\
    \approx & \frac{ \dot{N}_{s} }{ \sqrt{ 2 f_{s} } } \sqrt{ \Delta t} \quad (\dot{N}_{p} \gg f_{s}, \dot{N}_{D}, \dot{N}_{s}),
\end{split}
\end{equation}
where $\dot{N}_{i}$'s are the photon rates for $N_{i(=D,\, s,\, p)}$, and $\Delta t$ is the acquisition time.
If the sampling rate is smaller than the axion bandwidth, $\Delta f_{a}$, the ratio $f_{s} / \Delta f_{a}$ corresponds to the detection efficiency since integrating for more time does not increase the variance.
On the other hand, if the sampling rate is larger than $\Delta f_{a}$, the signals within the coherence time are correlated.
So, before calculating the variance, the number of photons within the axion coherence time interval should be summed up to avoid the correlation effects, limiting the maximum effective sampling rate to be the axion bandwidth.

\begin{figure}[tbp]
  \centering
  \includegraphics[width=0.99\linewidth]{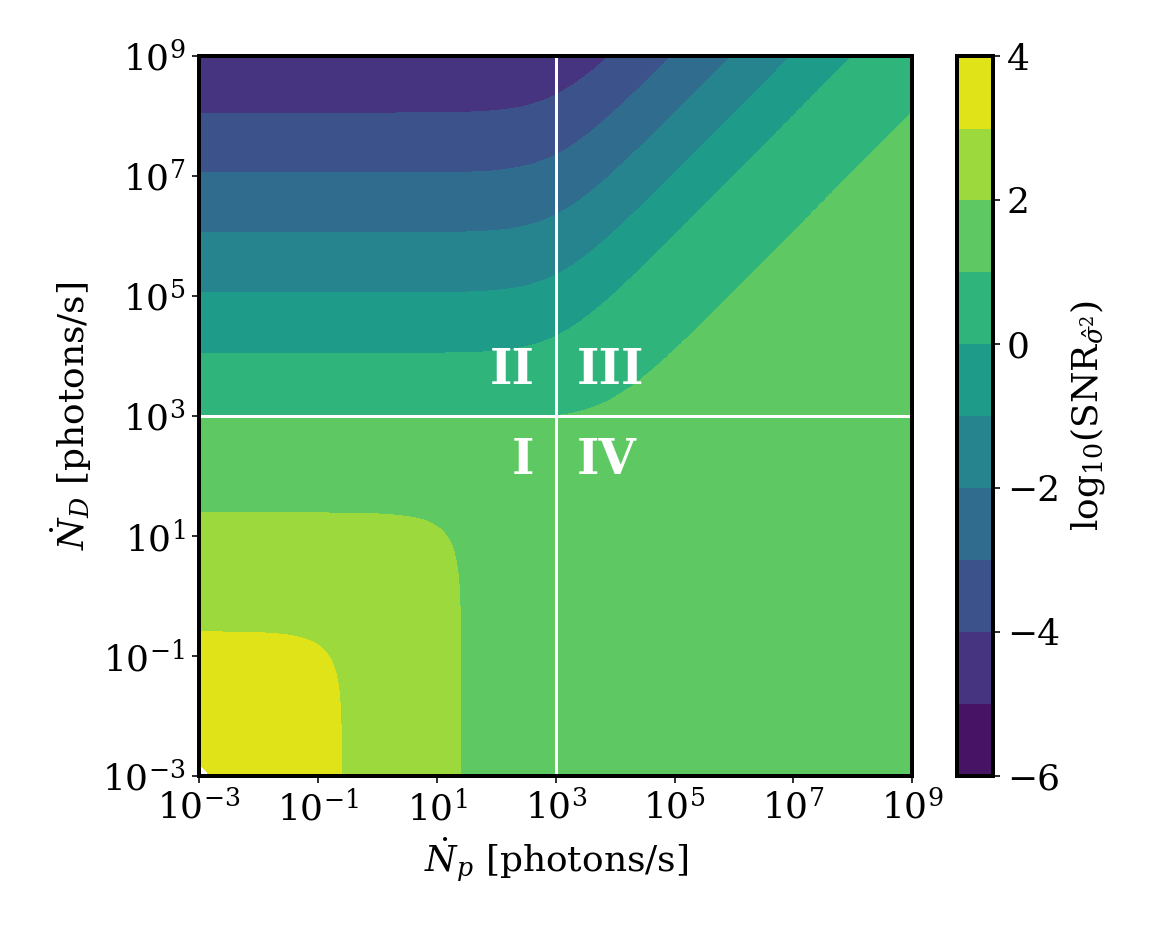}
  \caption{S/N defined in Eq.~\ref{eq:SNR_rate} for the dark count rate and the probe photon rate with target frequency of $1\,$GHz, sampling rate of $1\,$kHz, signal photon rate of $25\,$photons/s, and acquisition time of $400\,$s.
  The white solid lines are located at the sampling rate dividing the figure into four distinct regions.
  }\label{fig:snr}
\end{figure}

\Cref{fig:snr} shows the signal-to-noise ratio obtained from Eq.~\ref{eq:SNR_rate} as a function of dark count and probe photon rates.
The plot is divided into 4 regions by the sampling rate, ideally equal to the inverse of the axion coherence time.
When the dark count rate is less than the sampling rate, regions I and IV, injecting probe photons into the cavity reduces the signal-to-noise ratio.
On the other hand, if the dark count rate is larger than the sampling rate and more than one dark count is expected per sampling, regions II and III, the signal-to-noise ratio can be increased significantly by injecting probe photons, converging to the case when the dark count rate is equal to the sampling rate.

For example, when searching for an axion signal at $\SI{1}{GHz}$ and a quality factor of $10^{6}$, using a power detector with a high dark count rate designed to have a sampling rate of $\SI{1}{kHz}$, the effective dark count rate can be lowered to $\SI{1}{kHz}$ by injecting a sufficient number of probe photons.
Since the variance error is larger than the regular counting error by a factor of $\sqrt{2}$, the total acquisition time required for ${\rm S/N}=5$ is $80\,$s.
Haloscope experiments running at lower than 10 GHz, e.g., ADMX, CAPP, HAYSTAC, DMRadio~\cite{PRL2021ADMX,PRL2021Kwon,Nature2021Backes,arXiv2022DMRadio}, etc., have the most to gain with this new method.\footnote {If the axion signal is at $\SI{1}{MHz}$, then a  quality factor of $10^6$
would bring the integration (sampling) time to $\SI{1}{s}$. The amplified-variance method in that case brings the effective dark count rate to $\SI{1}{Hz}$, i.e., a thousand
times better than in the $\SI{1}{GHz}$ case.}

For comparison, the S/N obtained by a single photon detector, using the number of mean transmitted photons~\cite{PRD2013Lamoreaux}, is given by:
\begin{equation}
\label{eq:SNR_mu}
    {\rm S/N}_{\hat{\mu}} \approx \frac{ N_{s} }{ \sqrt{ N_{\rm th.} + N_{D} } } \sqrt{n} = \frac{ \dot{N}_{s} }{ \sqrt{ \dot{N}_{\rm th.} + \dot{N}_{D} } } \sqrt{\Delta t}.
\end{equation}
Comparing with Eq.~\ref{eq:SNR_rate}, the only difference is that the square root of the denominator changes from $2 f_{s}$ to $\dot{N}_{\rm th.} + \dot{N}_{D}$.
In other words, in the case that the dark count rate is higher than twice the sampling rate, the signal-to-noise ratio obtained by the amplified-variance method can be better than the conventional method.
For example, in an axion haloscope at $\SI{1}{GHz}$, a detector with a dark count rate higher than $\SI{2}{kHz}$, corresponding to 2 dark counts per ms, can achieve better sensitivity when using the amplified-variance method.

Furthermore, using the beating effect and a power detector with a short sampling time, it is also possible to be sensitive to axions with a frequency close but not exactly at the same frequency as the probing photons.
The beating effect due to the interference in terms of the number of transmitted photons is given by:
\begin{equation}
    2 \sqrt{\mathcal{N}_{p} \mathcal{N}_{a}} \cos\left[(\omega_{p} - \omega_{a})t + \phi \right],
\end{equation}
where $\mathcal{N}_{p,a}$ are the numbers of corresponding photons, $\omega_{p,a}$ are the angular frequencies of the pump and the axion signals, respectively, and $\phi$ is the phase of the axion signal relative to the pump.
It is known that $\mathcal{N}_{p,a}$ follows the Poisson distribution for each sampling and $\phi$ approximately follows the uniform distribution for each coherence time.
This corresponds to $2 \sqrt{\mathcal{N}_{p} \mathcal{N}_{a}} \cos\phi$ of the real part of the Fourier component at the beating frequency.
The Fourier component at the beating frequency also oscillates at each coherent time.
Therefore, the variance in the Fourier component has the same sensitivity at the sideband frequencies.

By rearranging \Cref{eq:SNR_rate} and assuming a detector with sufficiently short sampling time, the scanning rate in the haloscope experiments is:
\begin{equation}
    \frac{df}{dt} \approx \frac{\Delta f_{c}}{\Delta t} 
    = \frac{ \dot{N}_{s}^{2} }{ {\rm SNR}_{\hat{\sigma}^{2}}^{2} } \frac{ \Delta f_{c} } { 2 \Delta f_{a} } \quad (\dot{N}_{p} \gg \Delta f_{a}, \dot{N}_{D}, \dot{N}_{s}),
\end{equation}
where $\Delta f_{c}$ is the cavity bandwidth, and the effective sampling rate is set to match the axion bandwidth, $\Delta f_{a}$.

\begin{figure}[tbp]
  \centering
  \includegraphics[width=0.99\linewidth]{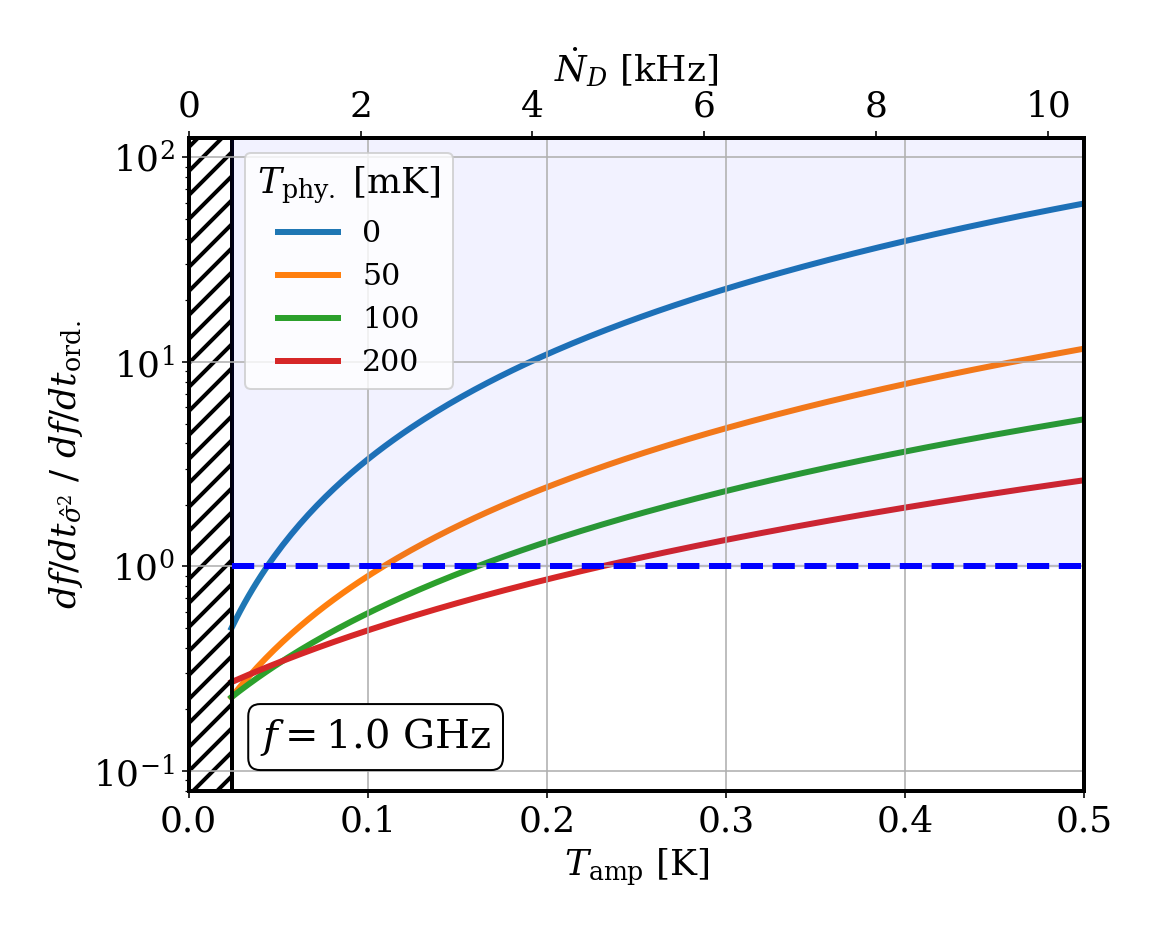} \\
  \caption{The scan rate enhancement compared to the ordinary acquisition method versus the noise temperature of the amplifier (added noise) under various physical cavity temperature conditions.
  The search frequency here is assumed to be $\SI{1}{GHz}$.
  The equivalent dark count rate on the top axis is derived by assuming the detector bandwidth to be the axion bandwidth.
  The blue shaded region represents where the performance of the amplified variance method exceeds that of the ordinary method.
  The hatched area is where the amplifier noise would be less than the quantum limit, and hence impossible for linear amplifiers without squeezing.
  }\label{fig:snr_improve}
\end{figure}

To determine the increase in scanning rate with the newly proposed detection method, we compared it with the ordinary method~\cite{JCAP2020Kim}. %for a system with the LTS-$\SI{12}{T} / \SI{320}{mm}$ magnet of Oxford Instruments at CAPP.
In Figure~\ref{fig:snr_improve}, the scanning rate using the amplified-variance method was compared with the scanning rate using the ordinary method when the axion haloscope was performed around $\SI{1}{GHz}$ under various physical cavity temperatures.
When the amplifier noise was close enough to the quantum limit, the traditional scanning rate was faster, due to the loss of a factor of square root of 2 when applying the variance estimator.
On the other hand, once the amplifier noise is sufficiently large compared to the quantum limit, the variance method can be expected to improve the scanning rate significantly.

\section{Discussion}

The method can also be used to reveal the location of hidden signals in the spectrum, when their variance characteristics are not random.
As an example, detectors such as bolometers have little preference for input frequency, but have a high noise equivalent power (NEP) (corresponding to the dark count) and have a large bandwidth.
However, if the variance detection method is used, it is possible to target only a specific frequency, and at the same time, the effective noise in terms of dark count rate can be reduced to the axion bandwidth.

As with the optical heterodyne method, the stability of the probe source used greatly affects the experimental sensitivity.
A stable probe source must be used, and more to that effect, the probe power can be branched early to correct its fluctuations, in order to reduce the related errors.
In particular, if a squeezed light source is used to provide the probe photons, the limiting statistical uncertainty can be further reduced~\cite{PT2011Miller} significantly.

Since this method basically uses a variance estimator, it has a disadvantage by a factor of $\sqrt{2}$ in S/N compared to a single photon detection method.
Therefore, there is no need to use this method when the sensitivity of a power detector itself is better than the standard quantum noise limit.
However, in the sub-MHz to tens of GHz region, using the amplified-variance estimator can be effective because most sensitive bolometer detectors still have a high dark count rate.

\section{Conclusions}

We have demonstrated analytically and numerically that probing the amplified-variance content of each frequency bin may be more advantageous than looking at its average power content.
Using the variance method compensates for the lack of knowledge of the axion phase.
Even if the dark count rate of the power detector is high, a near-quantum-noise limit can always be obtained by setting up a precise heterodyne interferometer.
The variance estimator for each known coherence time with a stable external probe signal gives an effective dark count rate of twice the sampling rate, i.e., two photon quanta per sampling.
Moreover, if the sampling rate of the detector is shorter than the coherence time, sideband signals can also be searched, using the beating effect.
In the last case, almost all axion dark matter experiments currently in progress should find the suggested method most advantageous over the traditional one.

\appendix

\section{Axion-induced power variance enhancement}
\label{sec:var_enhance}

If the axion frequency and phase were known, we could have injected external power~\cite{arXiv2021Sikivie} at the same frequency and $\pi/2$ relative phase to enhance the axion production rate.
Then the cavity detector would qualitatively behave similarly to a simple driven harmonic oscillator.
\begin{equation}
    \frac{d^2 X}{dt^2} + \gamma \frac{dX}{dt} +\omega^2_0 X = E_d \cos{(\omega t)},
\end{equation}
with the power decay lifetime $\tau = 1 / \gamma$ being the time it takes for the power to be reduced to $1/e$ of its initial value.
The cavity quality factor is given by $Q = \omega_0 / \gamma$.
$X$ is the axion field in arbitrary units.
The general solution is given by,
\begin{eqnarray}
   X(t) &=& A_1 e^{-\gamma t/2} \cos{(\sqrt{(\omega_0^2 - \gamma^2/4)}\, t + \phi_1)} + \nonumber \\
  & & E_d \frac{\omega \gamma \sin{\omega t} +(\omega_0^2 - \omega^2) \cos{(\omega t)}}{(\omega_0^2 - \omega^2)^2 + \omega^2 \gamma^2},
  \label{eq:resonance}
\end{eqnarray}
with the first term ignored for large times $\gamma t\gg 1$, while the second term consists of a cosine and a sine term. The power injected is proportional to the velocity times the force, i.e., $\frac{dX}{dt} E_d \cos{(\omega t)}$, with the average power given by
\begin{eqnarray}
    \langle P \rangle &=& \langle \frac{dX}{dt} E_d  \cos{(\omega t)} \rangle \nonumber \\
    &=& \frac{1}{2} E_d^2  \frac{\omega^2 \gamma } {(\omega_0^2 - \omega^2)^2 + \omega^2 \gamma^2},
\end{eqnarray}
since only the sine term from Eq.~(\ref{eq:resonance}) survives the integration. At resonance, it simplifies to
\begin{equation}
    \langle P \rangle = \frac{1}{2} E_d^2  \frac{Q } {\omega_0}.
\end{equation}

Clearly, when the external driving term is a cosine, the oscillating term phase in the cavity gets shifted by $\pi/2$. If we want this term to be accumulating on the axion driven oscillation, it would then have to be also $\pi/2$ off the axion phase with the correct sign.

The total axion plus the external driving term will accumulate a total power of
\[P_T = \left[E_d \sin{(\omega_d t + \phi_d)}+ E_a \cos{(\omega_a t + \phi_a)}\right]^2.\] 
When the driving frequency is the same as the axion frequency, then the total power is equal to,
\begin{equation}
\begin{split}
  P_T =& (1/2) E_d^2  + (1/2) E_a^2 \\
  & + 2 (E_d \sin{(\omega_a t + \phi_d)})(E_a \cos{(\omega_a t + \phi_a)}).  
\end{split}
\end{equation}
The oscillating term is equal to zero if one integrates for a longer time than the axion coherence time, estimated~\cite{arXiv2021Sikivie} to be of order of  $\SI{1}{ms}$ for $\SI{1}{GHz}$ axions. However, the characteristic effect is to greatly enhance the variance of the power of the particular frequency bin when we sample it within a time shorter than the expected axion coherence time. To ensure that the variance will be large, even in the case where the axion has a really long coherence time, one can randomly select the phase of the driving term.

\section{Test bench measurements}
\label{sec:test}

A pilot-experiment  at room temperature
% Zh.: "pseudo-experiment" sounds bad it was a real measurement, maybe call it "pilot (or test) room temperature experiment" 
was performed to test whether the variance increases when injecting a probe in the presence of the signal as expected.
The signal from a network analyzer was used as the probe, and a random phase axion signal was generated from the signal generator.
The signal and the probe were synthesized with a directional coupler just before injection into the cavity.
The network analyzer scanned the vicinity of the resonance frequency of the test cavity, and the transmission was repeatedly measured with an axion signal at the resonance frequency each time.
Although this experiment is a double quadrature measurement using a network analyzer, when enough iterations are performed with a power meter, it is classically the same as when only power is used in double quadrature measurement. %Perhaps these 2 para warrant a simple illustration

\begin{figure*}
\hfill
\subfigure[Frequency vs. power]{\includegraphics[width=8cm]{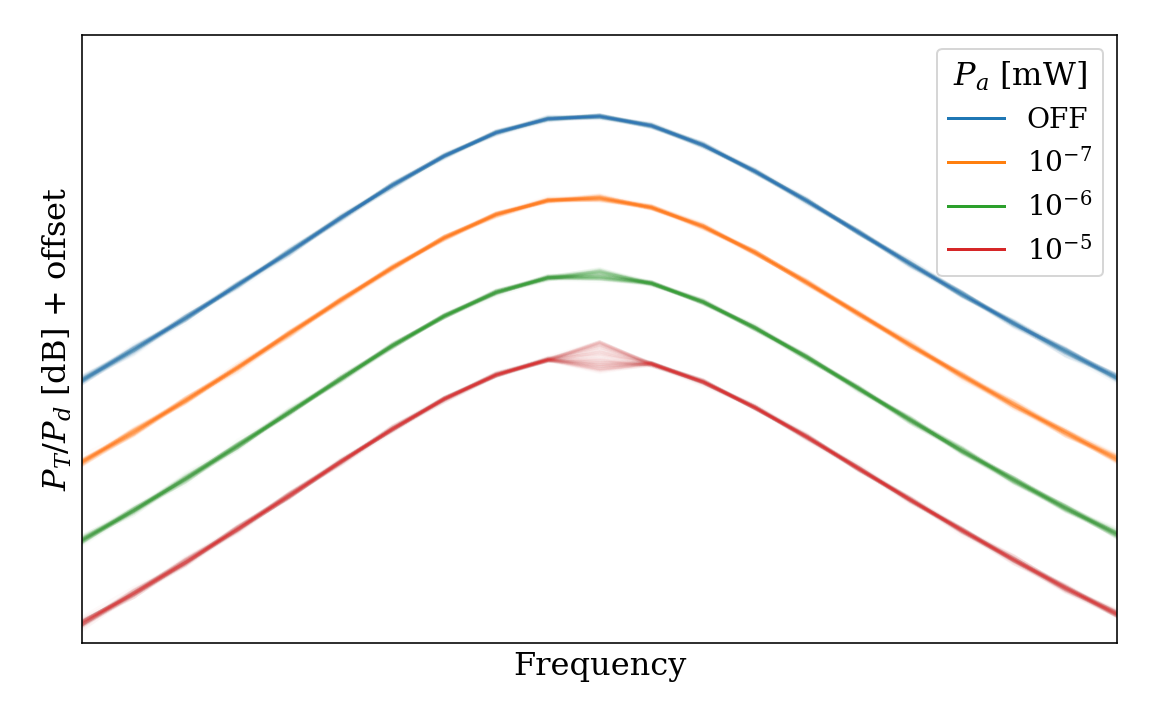}}
\hfill
\subfigure[Histograms of the peak power]{\includegraphics[width=8cm]{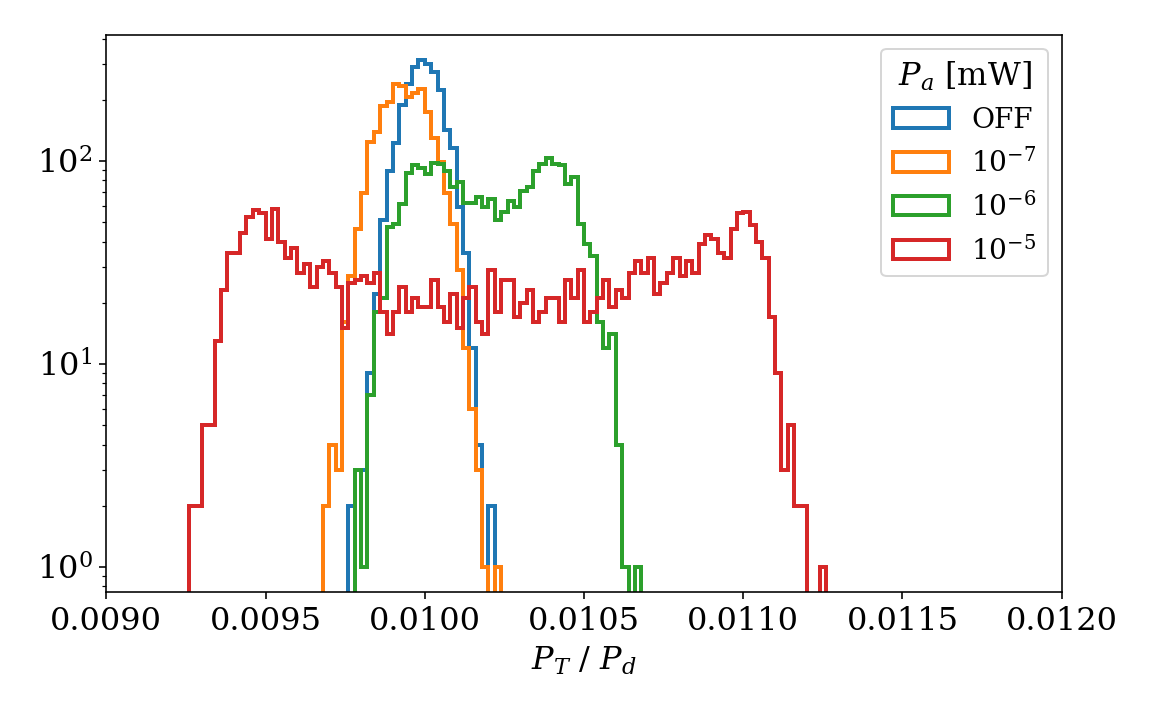}}
\hfill
\caption{(a) Power transmission vs. frequency at a constant probe power and increasing axion signal power that is indicated by the color of the trace. Arbitrary offsets are added to visually separate the traces. Notably, for the axion power of $\SI{e-5}{mW}$  (in red), the measured power at the resonant frequency (middle) fluctuates. Such fluctuation of the power can be measured to indicate the underlying axion power, since the probe power is known a priory. Distributions of the measured power at the peaks of the traces are noted and plotted separately --- (b). \\
(b) The histograms show the distribution of the peak powers of (a) with matching colors. In the presence of axion signal power (orange, green, and red) traces a typical sine-wave histogram distribution is observed. Such significance of ``sine-waveness'' in the distributions can be quantified to report the underlying axion power. Such sine-wave like distribution of the peak powers from (a) suggest that the measurement of the aforementioned probe-signal interference is measurable.}\label{fig:transmission_for_Ps}
\end{figure*}

\begin{figure}[t]
\centering
  \includegraphics[width=0.99\linewidth]{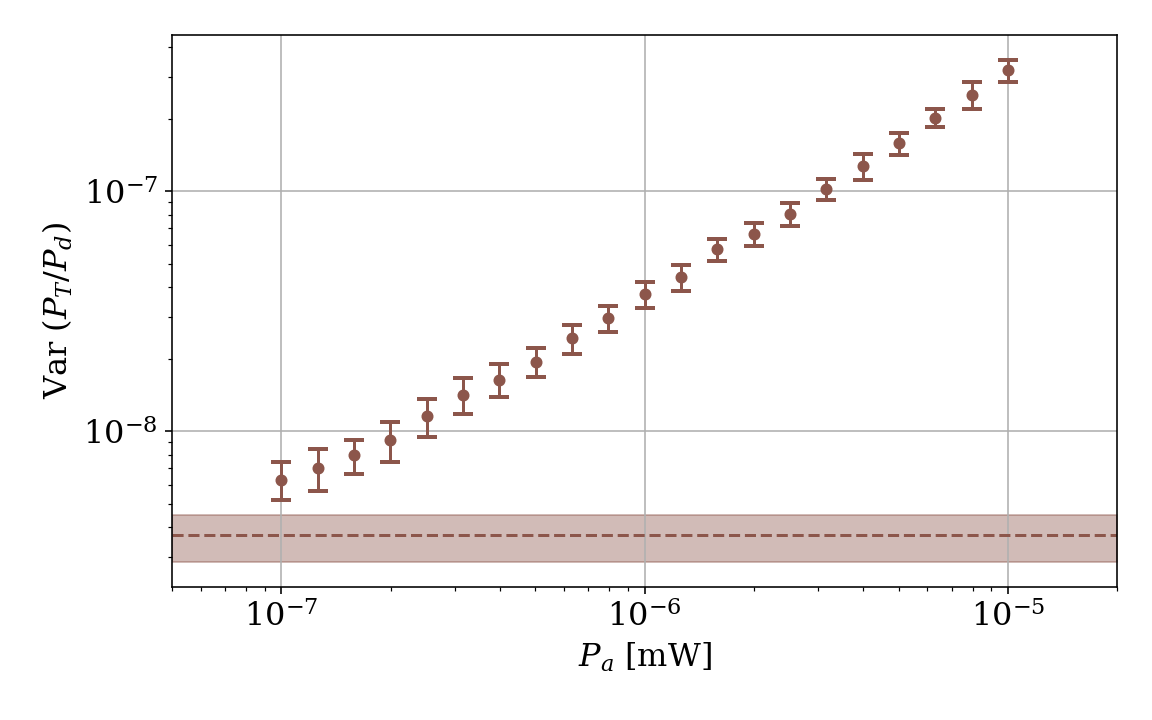}
  \caption{The recorded variance as a function of the applied signal power to be detected.
  The marker represents the average of the variance measurement and the error-bar refers to the error of the variance measurement with 50 traces.
  The dashed line is the case when there is no signal power with the shaded area as the deviation of it.
  }\label{fig:variance_vs_Ps}
\end{figure}

The results are illustrated on Fig.~\ref{fig:transmission_for_Ps} showing (a) multiple traces of power spectra with increasing axion power at a constant probe power and (b) the observed power distribution at the peaks of the curves (axion resonance frequency).
% This is best illustrated by an example --- \Cref{fig:transmission_for_Ps} (a) shows multiple traces of power spectra with increasing axion power at a constant probe power. The observed power distribution at the peaks of curves (axion resonance freq.) are shown separately --- \Cref{fig:transmission_for_Ps} (b). Such 
A clear variance was observed at the resonance frequency that we injected in, and the histogram shows that the peaks appear at both ends for sufficient signal power, indicating that it follows the distribution of a sine function.
As the strength of the axion signal increases, the dispersion increases while the average remains almost the same.
The slight change in the mean is presumed to be due to gain drift of the amplifier located between the cavity and the network analyzer.
Fig.~\ref{fig:variance_vs_Ps} shows the variance versus the signal power including statistical errors.
As expected from Eq.~\ref{eq:variance_photon}, the variance has very good linearity with the signal power.

\section{Signal-to-noise ratio}\label{sec:snr}
The signal-to-noise ratio of the variance-based method is as follows by definition:
\begin{equation}
    {\rm SNR}_{\hat{\sigma}^{2}} \equiv \frac{ \hat{\sigma}^{2}(N_{T}) - \hat{\sigma}^{2}(N_{T})|_{N_{s} = 0}}{ \sqrt{ \hat{\sigma}^{2} (\hat{\sigma}^{2}(N_{T})|_{N_{s} = 0}) } }.
\end{equation}
The term in the numerator is readily available from Eq.~\ref{eq:variance_photon} while the mean squared error of the variance estimator in the denominator can be obtained with the fourth moment ($\mu_{4}$) and variance ($\sigma^{2}$) of the distribution of the transmitted photons at the detector.
\begin{equation}
    \hat{\sigma}^{2}(\hat{\sigma}^{2}( N_{T} ) ) = \frac{\mu_{4} - \sigma^{4} (n - 3) / (n - 1)}{n} \approx \frac{\mu_{4} - \sigma^{4}}{n},
\end{equation}
where
\begin{equation}
\begin{split}
    \mu_{4} &= \langle N_{T}^{4} \rangle -4 \langle N_{T} \rangle \langle N_{T}^{3} \rangle + 6 \langle N_{T} \rangle^{2} \langle N_{T}^{2} \rangle -3 \langle N_{T} \rangle^{4}. \\
    \sigma^{2} &= \langle N_{T}^{2} \rangle - \langle N_{T} \rangle ^{2},
\end{split}
\end{equation}

The distribution of the transmitted photons follows:
\begin{equation}
    N_{T}|_{N_{s} = 0} \sim \mathcal{N}_{\rm th.} + \mathcal{N}_{D} + \mathcal{N}_{p} + 2 \sqrt{\mathcal{N}_{p} \mathcal{N}_{\rm th.}} \cos{\phi},
\end{equation}
where $\mathcal{N}_{\rm th.}$, $\mathcal{N}_{D}$, $\mathcal{N}_{p}$ follow Poisson distributions with the expected rates of $N_{\rm th.}$, $N_{D}$, $N_{p}$, respectively, and $\phi$ follows a uniform distribution between $-\pi$ and $\pi$.
Therefore, from the given distribution, the signal-to-noise ratio is:
\begin{equation}
\label{eq:snr_general}
    \begin{split}
        {\rm SNR}_{\hat{\sigma}^{2}} \approx& \frac{N_{s} + N_{s} N_{p}}{ \sqrt{\mu_{4} - \sigma^{4} }} \sqrt{n}, \\
        =& \frac{N_{s} + N_{s} N_{p}}{ \sqrt{C_{1} + C_{2} N_{\rm th.} + C_{3} N_{\rm th.}^{2}}  } \sqrt{n},
    \end{split}
\end{equation}
with
\begin{equation}
\begin{split}
    C_{1} &= (N_{D} + N_{p})^{2} ( 2 + 1 / (N_{D} + N_{p}) ), \\
    C_{2} &= 1 + 4 N_{D} + 2 N_{p} ( 5 + 7 N_{p} + 4 N_{D} ), \\
    C_{3} &= 2 ( 1 + 7 N_{p} + N_{p}^{2} ), \\
\end{split}
\end{equation}
When $N_{\rm th.} \to 0$, Eq.~\ref{eq:snr_general} becomes Eq.~\ref{eq:snr}.
By replacing the number of photons with the photon rates divided by the sampling rate, the signal-to-noise ratio in terms of the photon rates is:
\begin{equation}
    {\rm SNR}_{\hat{\sigma}^{2}} \approx \frac{\dot{N}_{s} + \dot{N}_{s} \dot{N}_{p} / f_{s} }{ \sqrt{\mathbb{C}_{1} + \mathbb{C}_{2} \dot{N}_{\rm th.} + \mathbb{C}_{3} \dot{N}_{\rm th.}^{2} }  } \sqrt{f_{s} \Delta t},
\end{equation}
with
\begin{equation}
\begin{split}
    \mathbb{C}_{1} &= (\dot{N}_{D} + \dot{N}_{p})^{2} ( 2 + f_{s} / (\dot{N}_{D} + \dot{N}_{p}) ), \\
    \mathbb{C}_{2} &= f_{s} + 4 \dot{N}_{D} + 2 \dot{N}_{p} ( 5 + 7 \dot{N}_{p} / f_{s} + 4 \dot{N}_{D} / f_{s} ), \\
    \mathbb{C}_{3} &= 2 ( 1 + 7 \dot{N}_{p} / f_{s} + (\dot{N}_{p} / f_{s})^{2} ), \\
\end{split}
\end{equation}
For a sufficient amount of probe photon rate, $\dot{N}_{p} \gg \dot{N}_{D}$ and $\dot{N}_{p} \gg f_{s}$, the signal-to-noise ratio is further reduced.
\begin{equation}
    {\rm SNR}_{\hat{\sigma}^{2}} \approx \frac{ \dot{N}_{s} }{ \sqrt{f_{s}^{2} + 7 \dot{N}_{\rm th.}f_{s} + \dot{N}_{\rm th.}^{2} } } \sqrt{\frac{f_{s} \Delta t}{2}}.
\end{equation}

% Please do not change these lines, keep the bibliography sepaparate in the
% 'main.bib' file (zhanibek)
\bibliography{main}
% \bibliographystyle{unsrt}
% \begin{thebibliography}{9}
% \bibitem{PRL1977PQ}
% \bibitem{PRL1978Weinberg}
% \bibitem{PRL1978Wilczek}
% \bibitem{PRL1979Kim}
% \bibitem{PRL1979Kim}
% \bibitem{article:Turner}
% \bibitem{NPB1980SVZ}
% \bibitem{YF1980Zhitnitsky}
% \bibitem{PLB1981DFS}
% \bibitem{PLB1983Wilczek}
% \bibitem{PLB1983Abbott}
% \bibitem{PLB1983Dine}
% \bibitem{PRL1983Sikivie}
% \bibitem{PRD1985Sikivie}
% \bibitem{JCAP2020Kim}
% \bibitem{arXiv2021Sikivie}
% \bibitem{RSI1946Dicke}
% \bibitem{NPhysics2012HoEom}
% \bibitem{PRL2017Brubaker}
% \bibitem{PRXQ2021Malnou}
% \bibitem{PRL2021ADMX}
% \bibitem{SupercondSci2021Kutlu}
% \bibitem{Nature2021Backes}
% \bibitem{SciAdv2022Yannis}
% \bibitem{PRD2013Lamoreaux}
% \end{thebibliography}

\end{document}